\documentstyle [12pt]{article}
\newcommand {\sg} {\sigma}
\begin{document}
\baselineskip=20.0pt
\begin{titlepage}
\begin{flushright}
{\large {\bf UTCCP-P-80 } } \\
\end{flushright}
\vspace{1cm}

\begin{center}
{\Large {\bf Short-time dynamics and magnetic critical behavior}}\\
\vspace{0.2 cm}
{\Large {\bf of two-dimensional random-bond Potts model}}

\vspace{1.5 cm}

{\bf \large He-Ping Ying}\footnote{Work supported in part by the 
NNSF of China under Grant No. 19975041} 

\small {\it  Center for Computational Physcs, University of Tsukuba, Tsukuba, 
Ibaraki 305-8577, Japan, \\and Zhejiang Institute of Modern Physics, Zhejiang 
University, Hangzhou 310027, P.R. China~\\ }

\vspace{0.5 cm}
{\bf \large Kenji Harada}

\small {\it  Department of Applied Analysis and Complex Dynamical Systems,
Kyoto University,~~~~~~~~~~~\\ Kyoto 606-8501, Japan }

\vspace{1.0 cm} 
{\today}
\end{center}

\vspace{0.8cm}
\begin{abstract}
The critical behavior in the short-time dynamics for the random-bond Potts 
ferromagnet in two-dimensions is investigated by short-time dynamic
Monte Carlo simulations. The numerical calculations show that this dynamic 
approach can be applied efficiently to study the scaling characteristic, which 
is used to estimate the critical exponents $\theta$, $\beta/\nu$ and $z$,
for the quenched disorered systems from the power-law behavior of the 
$k$th moments of magnetizations.

\end{abstract}
\vspace{0.6cm}
\vspace{0.8cm}
 
\end{titlepage}
\section{Introduction}
The understanding of the role played by
quenched impurities on the nature of phase transitions is one
of the significant subjects in statistical physics, and it has been 
a topic of substantial interest for many authors in the last two decades
\cite{Imry79,Hui89,Aize89,Chen92,Chen95,Picco96,Cardy96,Cardy97,Chate98}.
According to the Harris criterion\cite{Har74}, quenched randomness
is a relevant perturbation at the second-order critical point when 
its specific-heat exponent $\alpha$ of the pure system is positive.
Following the earlier work of Imry and Wortis\cite{Imry79} who argued that 
a quenched disorder could produce rounding of a {\it first-order} phase 
transition and thus induce {\it second-order} phase transitions, the introduction
of randomness to systems undergoing a first-order phase transition is
comprehensively considered. It was shown by Hui and Berker that the bond 
randomness can have a drastic effect on the nature of a first-order phase 
transition by phenomenological renormalization group arguments \cite{Hui89},
and the feature has been placed on a firmer basis with the rigorous proof of 
vanishing of the latent heat \cite{Aize89}. Their theory was numerically checked 
with the Monte Carlo (MC) method by Chen, Ferrrngberg and Landau (CFL)
\cite{Chen92,Chen95}, where the eight-state Potts model  was studied with 
random-bond disorders.  Experimental evidence
in two-dimensional systems was found that in the order-disorder
phase transitions of absorbed atomic layers, the critical exponents
are modified, in the addition of disorder, from the original four-state
Potts model universality class in the pure case \cite{Sch94,Voges98}. 
On the other hand, no modification is found when the pure system 
belongs to the Ising Universality  class \cite{Mohan98}. 
The theoretical study of such kinds of disordered systems is also an active 
field where a resort to intensive MC
simulations is often helpful \cite{Chate99,Wise95,Kim96,Olson99,Kar95,Yasar98}.

It is well known that the pure Potts model in two-dimensions (2D) has a 
second order phase transition when the number of Potts state $q \le 4$
and is first order for $q > 4$. As the specific
heat exponent $\alpha$ of the pure system is always positive for 
$q>2$, the disorders will be the relevant perturbation for the 
Potts model. As a result, all
the transitions are second order for all 2D $q$-state Potts
models in the presence of quenched disorders, and the impurities 
have particularly strong effect for $q > 4$, even changing the 
order of the transitions. 

In this paper, we discuss the dynamic scaling features of random-bond Potts 
model (RBPM) through MC simulations, to estimate the critical 
exponents. We consider the important questions of whether there exists 
an Ising-like universality class for the RBPM and how is the critical 
behavior affected by the introduction of disorder into the pure 
system \cite{Olson99}.
The large-scale MC results by the CFL and in ref.\cite{Kar95} suggest that, in
2D, any random systems should belong to the pure Ising universality class. 
These results are also coherent with  experiment \cite{Sch94}.
In recent papers\cite{Cardy96,Cardy97}, however, Cardy and Jacobsen 
studied the random-bond Potts models for several values of $q$ with 
a different approach based on the connectivity transfer matrix (TM)
formalism of Bl\"ote and Nightingale \cite{Blote82}
Their estimates of the critical
exponents led to a continuous variation of $\beta/\nu$ with $q$, which
is in sharp disaggreement with the MC results for $q=8$
\cite{Chen92,Chen95}.
We hope that the resulting critical behavior measured in this paper will 
play a role in settling this controversy. 
Furthermore we will test the {\it short-time dynamic} (STD) MC 
approach for the first time, to study the spin systems with the
quenched disorder and to show its efficiency by numerical
studies, which is also one of our main aims of this paper.

\section{Model and Method}
The Hamiltonian of $q$-state Potts model with quenched random 
interactions can be written,
\begin{eqnarray}
-\beta H=\sum_{<i,j>} K_{ij} \delta_{\sg_i \sg_j}~,~~~~ K_{ij} > 0,~~~~~
\label{Ham1}
\end{eqnarray}
where the spin $\sg$ can take the values 1,$\cdots ~ q$, 
$\beta=1/k_B T$ the inverse temperature, $\delta$
is the Kronecker delta function, and the sum is over all 
nearest-neighbor pairs on a 2D lattice. The 
dimensionless couplings $K_{ij}$ are selected from two positive 
(ferromagnetic) values of $K_1$ and $K_2=rK_1$, with a strong 
to weak coupling ratio $r=K_2/K_1$ called as {\it disorder amplitude}, 
according to a bimodal distribution,
\begin{eqnarray}
P(K)=p\delta (K -K_1) + (1-p)\delta (K -K_2)~.
\label{eq2}
\end{eqnarray}
When $p=0.5$, the system is {\it self-dual} and the exact 
critical point can be determined by \cite{Kin81},
\begin{eqnarray}
( e^{K_{c}} - 1)(e^{K'_{c}} - 1)= q~.
\label{eq3}
\end{eqnarray}
where $K_{c}$ and $K'_{c}$ are the corresponding critical values of $K_1$ and 
$K_2$ respectively at the transition point. While $r=1$ corresponds to the pure 
case and the critical point is located at $K_c=\mbox{log}(1+\sqrt {q~})$ and  
the phase transitions are first-order for $q > 4$. 
With additional random-bond distribution, however, 
new second-order phase transitions are induced for any of $q$-state 
Potts models and the new critical points
are determined according to Eq.(\ref{eq3}) for different values of
disorder amplitude $r$ and state parameter $q$. 

In this work we chose $q=8$, which is known to have a strong  
first-order phase transition, in the hope that we would find a new 
second-order phase transition caused by the quenched disorder to show 
the  effect of impurities on the {\it first-order} systems.
The strength of the disorder was chosen for several values of $r$,
as was done in \cite{Chen95,Picco96,Chate99}, to check the Ising-like 
universality class.
To minimize the number of bond configurations needed for the
disorder averaging, we confined our study to the bond distributions in 
which there are the same number of strong and weak bonds in each of 
the two lattice directions. This procedure should reduce the variation 
between different bond configurations, with no loss of generality.

We performed our simulations by the STD 
method \cite{Zheng98} on the 2D square lattices with 
periodic boundary conditions. This dynamic MC simulations have 
been successfully performed to estimate the critical temperatures 
$T_c$ and the critical exponents $\theta$, $\beta$, $\nu$ and 
dynamic exponent $z$ for the 2D Ising model \cite{Li95} and the 
2D 3-state Potts model \cite{Okano97}, since for both models there
exist second order phase transitions.  Recently this approach 
has also been extensively applied for the Fully 
Frustruated XY model and spin glass systems to study
the critical scaling characteristics to estimate all the dynamic and 
static critical exponents \cite{Luo98,Luo99,Ying98}. 

Traditionally it was believed that universality and scaling relations
can be found only in the equilibrium stage or {\it long-time} regime. 
In Ref.\cite{Janss89},  however
it was discovered that for a magnetic system in states with a very high 
temperature $T \gg T_c$ which is suddenly quenched to the critical 
temperature $T_c$ and then evolve according to a dynamics of model A
\cite{Hoh77}, there emerges a universal dynamic scaling behavior 
already within the short-time regime, which satisfies, 
\begin{eqnarray}
M^{(k)}(t,\tau,L,m_0)=b^{-k\beta/\nu}M^{(k)}(b^{-z}t,b^{1/\nu}\tau,
b^{-1}L,b^{x_0}m_0), ~~~~~
\label{eq4}
\end{eqnarray}
where $M^{(k)}$ is the $k$th moment of the magnetization,
$\tau=(T-T_c)/T_c$ is the reduced temperature, $\beta$ and $\nu$ are
the well known static critical exponents and $b$ is a scaling factor.
The variable $x_0$, a {\it new independent} exponent, is the scaling 
dimension of initial magnetization $m_0$.  This dynamic scaling form is
generalized from finite size scaling
in the equilibrium stages\cite{Bind92}. Importantly
the scaling behavior of Eq.(\ref{eq4}) can be applied to both dynamic 
exponent measurements and the estimates of the static exponents 
originally defined in equilibrium.

We begin our study on the evolutions of magnetization in the
initial stage of the dynamic relaxation starting at very high
temperature and small magnetization ($m_0\sim 0$). For a sufficiently
large lattice ($L\rightarrow\infty$), from Eq.(\ref{eq4})
by setting $\tau=0$, $b = t^{1/z}$, it is easy to derive that
\begin{eqnarray}
 M^{(k)}(t,m_{0})=t^{-k\beta/\nu z}M^{(k)}(1,t^{x_{0}/z}m_{0}).~~~~
\label{eq5}
\end{eqnarray}
When $k=1$ we get the most important scaling relation on which
our measurements of the critical exponent $\theta$ are based,
\begin{eqnarray}
M(t) \sim m_0 t^\theta,~~~~~~~\theta = (x_0 - \beta/\nu)/z.~~~~ 
\label{cont1}
\end{eqnarray}
As a result, the magnetization undergoes an initial increase at the 
critical point $K_c$ after a microscopic time $t_{mic}$.
This prediction is supported by a number of MC 
investigations which have been applied to detect all the static and 
dynamic critical exponents \cite{Li95,Okano97} as well as the 
critical temperatures \cite{Luo98,Schue95}. 
The advantage of the dynamic MC simulations is that it may eliminate 
critical slowing down since the measurements 
are performed in the early time stages of the evolution 
where the spatial and time correlation lengths are small.  



In our simulations, the time evolution of $M(t)$ 
is calculated through the definition
\begin{eqnarray}
~~ M(t) &=& \frac{1}{N} [\frac{q <M_O> -1}{q-1}].
\label{eq6}
\end{eqnarray}
Here $M_O =\mbox{max}(M_1, M_2, \cdots, M_q)$ with $M_i$ being the 
number of spins in the $i$th state among $q$ states. 
$<\cdots>$ denotes 
the initial configuration averages over independent random number sequences,
and $[\cdots]$ the disorder configuration averages over quenched 
random-bond distributions. $N=L^2$ is a number of spins on this 
square lattice and $q=8$ is chosen. 

 
The susceptibility plays an important role in the equibrium. Its finite
size behavior is often used to determine the critical temperature
and the critical exponents $\gamma/\nu$ and $\beta/\nu$ \cite{Chen95}.
For the STD approach, the time-dependent susceptibility (the second
moment of the magnetization) is also interesting and important.
For the random-bond Potts model, the second moment of the magnetization
is  usually defined as
\begin{eqnarray}
~~ M^{(2)}(t) &=&  \frac{1}{N} [(<M^2(t)> - <M(t)>^2)]~.~~~~~~~~
\label{eq7}
\end{eqnarray}

To study the scaling behavior of the second moment of magnetization, 
we have to take the initial states of $m_0 =0$ to start the relaxation 
processes. Because the spatial correlation length in the beginning of 
the relaxation is small compared with the lattice size $L^d$
in the short-time regime of the dynamic evolition, the second moment 
behaves as $M^{(2)}(t,L)\sim L^{-d}$. Then the 
finite  size scaling Eq.(\ref{eq4}) induces a power-law 
behavior at the critical temperature, 
\begin{eqnarray}
 M^{(2)}(t) \sim t^{y},~~~~~~~~~~~~~~ y=(d-2\beta/\nu)/z.~~~~
\label{cont3}
\end{eqnarray}
From the scaling analysis of the spatial correlation function we easily 
realize the non-equibrium spatial correlation length 
$\xi \sim t^{1/z}$. Therefore $M^{(2)}(t) \sim \xi^{(d-2\beta/\nu)}$.

In the above considerations the dynamic relaxation process was assumed to start
from a disordered state or with small magnetization $m_0$. Another interesting
and important process is the dynamic relaxation from a completely ordered state.
The initial magnatization locate exactly at its fixed point $m_0=1$, where
scaling of the form,
\begin{eqnarray}
M^{(k)}(t,\tau,L)=b^{-k\beta/\nu}M^{(k)}(b^{-z}t,b^{1/\nu}\tau, b^{-1}L), ~~~~~
\label{eq8}
\end{eqnarray}
is  expected. This scaling form looks to be the same as the dynamic scaling one in 
the long-time regime, however, it is now assumed already valid in the macroscopic
short-time regime.

For the magnetization itself, $b=t^{1/z}$ yields, for a sufficiently large
lattice,
\begin{eqnarray}
M(t,\tau)=t^{-\beta/\nu z} M(1, t^{-\beta/\nu z} \tau). ~~~~~
\label{eq9}
\end{eqnarray}
This leads to a power-law decay behavior of
\begin{eqnarray}
M(t,\tau)=t^{-c_1} , ~~~~~c_1 =\beta/\nu z, ~~~~
\label{eq10}
\end{eqnarray}
at the critical point ($\tau =0$). The formula can be used to calculate 
the critical exponents $\beta/\nu$ and $z$. For a small but nonzero $\tau$, 
the power-law behavior will be modified by the scaling function
$ M(1, t^{-\beta/\nu z} \tau)$, which has been used to determine the 
critical temperatures \cite{Schue95,Jas99}. 
Furthermore, by introducing a Binder cumulant 
\begin{eqnarray}
U(t, L) = \frac{M^{(2)}(t,L)}{(M(t,L))^2} - 1~,~~~~
\label{eq11}
\end{eqnarray}
a similar power-law behavior at the critical point induced from the 
scaling Eq.(\ref{eq8}) shows that,
\begin{eqnarray}
 U(t, L) \sim t^{c_2}~, ~~~~~~~~c_2 = d/z~,
\label{eq12}
\end{eqnarray}
on a large enough lattice. 
Here, unlike the relaxation from the disordered state, the
fluctuations caused by the initial configurations are much smaller. 
In pratical simulations, these measurements of the critical exponents 
and critical temperature are better in quality than those from the 
realization process starting from disordered states.

\section{MC Simulations and Results }

\begin{table} \begin{center}
\caption {
The tendency and measured values of $\theta$ as a function of the disorder
amplitude $r$ for different initial $m_0$ at the critical points
$K_c$ on the lattice $64^2$  lattice.
          } 
\vskip 0.5cm
\begin{tabular}{ c c  c  c c | c | c } \hline\hline
$m_0 $  & 0.06     & 0.04     &  0.02    &  0.01   & $\theta$ &  $K_c(r)$    \\ \hline
$r= 2$  & 0.310(8) & 0.338(8) & 0.350(8) &0.352(7) & 0.353(6) & 0.920185271...\\ \hline
$r= 5$  & 0.160(6) & 0.215(6) & 0.252(5) &0.257(4) & 0.262(4) & 0.512307010...\\ \hline
$r= 8$  & 0.106(5) & 0.167(4) & 0.208(4) &0.218(3) & 0.221(3) & 0.367963156... \\ \hline
$r=10$  & 0.090(4) & 0.146(3) & 0.193(3) &0.202(3) & 0.203(3) & 0.312655667... \\ \hline\hline
\end{tabular}
\end{center} \end{table}
As it was pointed out that the Heat-bath algorithm is more efficient than 
the Metropolis algorithm in the STD \cite{Okano97}, and the
universality is satisfied for different algorithms, we only
perform the MC simulations with the Heat-bath algorithm at the
critical points of 2D eight-state RBPM for an optimal
disorder amplitude $r^* = 10$ which is located on the random fixed
point regime with a largest value of central charge $c=1.5300(5)$
\cite{Chate99}. Samples for averages are taken both for over 300 disorder
distribution configurations and about $\sim$500 independent
initial configurations on the square lattices $L^2$ with $L$ up 
to 128.  Statistical errors are simply estimated by performing
three groups of averages with
different random seed selectes for the initial configurations.
It should be noted that, except for $M(t)$, the measurements of $M^{(2)}(t)$ 
and $U(t)$ are restricted to the initial states with $m_0=0$ or $m_0=1$.
Importantly, it was verified that the critical exponents save the same value 
the same as those in the equilibrium or {\it long-time} stage of the 
relaxation \cite{Li95}. Therefore we can measure these exponents based
on the corresponding scaling relation in the initial stages of the
relaxation.

We start our simulations to verify the power-law scaling behavior
of $M(t)$ with several values of disorder
amplitude at the critical points $K_c(r)$ (as shown in Table 1).   
The initial configurations are prepared  with small magnetizations
$m_0 = 0.06, 0.04$, 0.02 and exact zero states.
In Fig.1, the time evolutions of the magnetization $M(t)$ versus the 
disorder amplitude $r$ on a $64^2$ lattice are 
displayed with a double-log scale. We can easily find that all 
the curves exhibit the power-law behavior predicted by 
Eq.(\ref{cont1}). Thus $\theta$ can be estimated from the slopes 
of the curves. The values of $\theta$ as a function of 
the disorder amplitude $r$ for small initial magnetization 
$m_0$ are presented in the Table 1. 
\begin{table} \begin{center}
\caption {
The values of scaling exponents for the 2D $q$=8 RBPM with $r=10$,
measured from the scaling functions of $M(t)$, $M^{(2)}(t)$ and $U(t)$
respectively starting from both the random initial states and ordered states.
Also listed are those for the 2D Ising and $q=3$ Potts models, and the 3D Ising 
model \cite{Zheng98,Okano97,Jas99}. } 
\vskip 0.5cm
\begin{tabular}{c | c | c  c  c | c } \hline\hline
     exponent         & $m_0$    & 2D RBPM  & 2D Ising  &  2D Potts  & 3D Ising \\ \hline
   $\theta$           & $\sim0.0$& 0.197(4)  & 0.191(1)  & 0.075(3)  & 0.108(2) \\ 
 $y=(d-2\beta/\nu)/z$ & ~~       & 0.438(6)  & 0.817(7)  & 0.788(1)  &  ~       \\ \hline
  $c_1=\beta/\nu z$   & $=1.0$   &  0.0390(6) & 0.056(1)  & 0.065(1) & 0.2533(7) \\
  $c_2= d/z$          & ~~       & 0.518(9)  & 0.926(8)  & 0.934(9) & 0.1462(12) \\ \hline
  $2\beta/\nu=d-yz$   & ~~       & 0.302(6)  & 0.240(36) & 0.269(7)  & 1.034(4) \\ 
$2\beta/\nu$ (exact)  &  ~       & ~~        &    1/4    &   4/15    &  ~    \\ \hline\hline
\end{tabular}
\end{center} \end{table}

We then set $m_0=0$ to measure the second moment of magnetization.  
The power-law behaviour of the second moment $M^{(2)}(t)$ is observed 
in Fig. 2, where the curves for different lattice sizes are plotted. 
Again, they present a very nice power-law increase. Values of scaling 
dimension $y=(d-2\beta/\nu)/z$ determined from slopes of the curves 
during $t =[10,200]$ are listed in the Table 2. 

We furthermore set $m_0=1$ to observe the evolution of the magnetization and 
the Binder cumulant, both should show the power-law behavior as predicted 
by the Eq.(\ref{eq10}) and Eq.(\ref{eq11}). Their curves are plotted in  the 
Figs. 3 and 4 respectively. The values of the scaling dimension $c_1=\beta/\nu z$ 
and $c_2=d/z$ are then estimated, as are also presented in Table 2. 
Now the results of $y$, $c_1=\beta/\nu z$ and $c_2=d/z$  can be
used to estimate critical exponent $\beta/\nu$ shown in Table 2.
For comparison, also listed in Table 2 are the correspoinding results 
of the scaling dimension for the Ising and $q=3$ Potts models on 
2D (3D) square (cubic) lattices, and in Table 3 we summarize the results of 
the critical exponent $\beta/\nu$ up to the present.


\begin{table} \begin{center}
\caption {
The results of magnetic scaling exponent $\beta/\nu$ estimated by different 
methods for the 2D eight-state RBPM. }
\vskip 0.5cm
\begin{tabular}{c| c| c |c } \hline\hline
      Authors      & $r$   &  $\beta/\nu$ & Technique \\ \hline
    CFL\cite{Chen92}             & ~~2~~~ & 0.118(2)  & MC \\ \hline
Cardy and Jacobsen\cite{Cardy97} & ~~2~~~ & 0.142(4)  & TM \\ \hline
Chatelain and Berche\cite{Chate98} & ~~10~~ & 0.153(3) & MC  \\ \hline
Picco\cite{Picco98}              & ~~10~~ & 0.153(1) & MC  \\ \hline
Present work                     & ~~10~~ & 0.151(3) & STD \\ \hline\hline
\end{tabular}
\end{center} \end{table}

\section{Summary and Conclusion}

In this paper we have investigated the short-time critical dynamics of 
the random-bond Potts model on 2D lattices to verify whether it has 
a second order phase transition in the Ising-like universality 
class, by a STD study. Dynamic scaling haviour 
was found, and has been used to estimate the critical 
exponents $\theta$, $z$ and $\beta/\nu$. 
Our main results are summarized in Table 2, which are 
obtained from the slopes of power-law curves of $M(t)$, $M^{(2)}(t)$ 
and $U(t)$ in the double-log scales by least--square fits.

Our work shows that for the RBPM, there exists a  
power-law behavior, which is the typical feature of a continuous phase 
transition in the STD processes. The $r$-dependence characteristics 
of the dynamic exponent $\theta$ gives evidence that their 
dynamic MC behavior is different from the pure Ising model, and  
the values of magnetic exponent $\beta/\nu$ in our calculation seems  
to be the same as that given by the TM formula \cite{Cardy97}, 
but not as that by the CFL \cite{Chen92,Chen95}. Furthermore we found that
the values of dynamic exponent $\theta$ depend on the disorder amplitude 
$r$, and it suggests that the dynamic exponent $z$ may also depend on 
the $r$, which would be interesting to examine for future study. 

In conclusion, this study presents numerical evidence that the quenched
impurities in the RBPM can induce new second-order phase transitions, but
they appear not always to belong to Ising-like universality class,
although the result of new critical exponent $\theta$ is the same for
both the $r=10$ RBPM and the Ising model within errorbars by present
calculations. Second, as the effect of critical slowing down in the equilibrium
stage for the RBPM is more severe than that for the pure systems,
the cluster algorithms, up to now, have been frequently used in MC simulations
of the 2D RBPM \cite{Chen92,Chen95,Chate98,Chate99,Yasar98}. In present paper,
however, we have applied the STD MC simulations which use local updating
schemes for the 2D RBPM, and the fact that dynamic MC simulations
can avoid the critical slowing down in the STD  processes, where the spatial
correlation length is still small, makes it easier to calculate the critical
exponents.
An important subject for further study is, for example, how the dependence
of dynamic critical exponent $\beta/\nu$ on the state parameter $q$ and
the disorder amplitude $r$ by a systematic simulation using the STD method
in order to clarify the crossover behaviour from the random fixed point to
a percolation-like limit \cite{Picco98}. This is being studied at present.


We acknowledge the helpful discussions with Y. Aoki and H. Shanahan. 
This research was initiated during the visit to University of Tsukuba 
(HPY, who also acknowledge the Center for Computational Physics for
hospitality where the MC simulations were performed on the DEC 
workstations). 



\vskip 2.0cm

\noindent FIGURE CAPTIONS

\vskip 0.9cm
FIGURE 1: 
The time evolution of magnetization showing the $r$--dependence of $\theta$,
plotted with a double-log scale on a lattice of $64 \times 64$ with $m_0=0.01$. 
 
\vskip 0.6cm
FIGURE 2: 
The time evolution of second moment of magnetization starting from absolute 
random states, plotted with a double-log scale on lattices of 
$32 \times 32$, $64 \times 64$ and $128 \times 128$. 

\vskip 0.6cm
FIGURE 3: 
The power-law decay of magnetization starting from fully ordered states, 
plotted with a double-log scale on lattices of $32 \times 32$, 
$64 \times 64$ and $128 \times 128$. The finite size effect is obvious
when the lattice size $L < 64$.

\vskip 0.6cm
FIGURE 4: 
The  time evolution of Binder comulant starting from fully ordered states, 
plotted with a double-log scale on lattices of $32 \times 32$, 
$64 \times 64$ and $128 \times 128$. 

\baselineskip=12.0pt
\begin{thebibliography}{99}
\bibitem{Imry79}
Y. Imry and M. Wortis, Phys. Rev. {\bf B9}(1979)3580.
\bibitem{Hui89}
K. Hui and A.N. Berker, Phys. Rev. lett. {\bf 62}(1989)2507.
\bibitem{Aize89} 
M Aizenman and J. Wehr, Phys. Rev. lett. {\bf 62}(1989)2503.
\bibitem{Chen92}
S. Chen, A.F. Ferrenbrerg and D.P. Landau, 
Phys. Rev. lett. {\bf 69}(1992)1213.
\bibitem{Chen95}
S. Chen, A.F. Ferrenbrerg and D.P. Landau, Phys. Rev. {\bf E52}(1995)1377.
\bibitem{Picco96}
M. Picco, Phys. Rev. {\bf B54}(1996)14930; 
Phys. Rev. lett. {\bf 79}(1997)2998.
\bibitem{Cardy96}
J. Cardy, J. Phys. {\bf A29}(1996)1897.
\bibitem{Cardy97}
J. Cardy and J.L. Jacobsen, Phys. Rev. lett. {\bf 79}(1997)4063.
\bibitem{Chate98}
C. Chatelain and B. Berche, Phys. Rev. Lett. {\bf 80}(1998)1670.
\bibitem{Chate99}
C. Chatelain and B. Berche, Phys. Rev. {\bf E58}(1998)R6899; {\bf E60}(1999)3853.
\bibitem{Har74}
A.B. Harris,  J. Phys. {\bf C7}(1974)129.
\bibitem{Sch94}
L. Schwenger, K. Buddle, C. Voges and H. Pfn\"ur,
Phys. Rev. lett. {\bf 73}(1994)296.
\bibitem{Voges98}
C. Voges and H. Pfn\"ur, Phys. Rev. {\bf B57}(1998)3345.
\bibitem{Mohan98}
Ch. V. Mohan, H. Kronm\"uller and M. Kelsch, 
Phys. Rev. {\bf B57}(1998)2701.
\bibitem{Wise95}
S. Wiseman and E. Domany,  Phys. Rev. {\bf E51}(1995)3074.
\bibitem{Kim96}
J.-K. Kim, Phys. Rev. {\bf B53}(1996)3388.
\bibitem{Olson99}
T. Olson and A.P. Young, Phys. Rev. {\bf B60}(1999)3328, and 
references therein.
\bibitem{Kar95}
M. Kardar, A.L. Stella, G, Sartoni and B. Derrida, 
Phys. Rev. {\bf E52}(1995)R1269.
\bibitem{Yasar98}
F. Yasar, Y. G\"und\"uc, and T. Celik, Phys. Rev. {\bf E58}(1998)4210.
\bibitem{Blote82}
H.W.J. Bl\"ote and M.P. Nightingale, Physica  {\bf A112}(1982)405.
\bibitem{Kin81}
W. Kinzel and E. Domany, Phys. Rev. {\bf B23}(1981)3421.
\bibitem{Zheng98}
B. Zheng, Int. J. Mod. Phys. {\bf B12}(1998)1419.
\bibitem{Li95}
Z.B. Li,  L. Sch\"ulke, and B. Zheng, Phys. Rev. Lett. {\bf 74}(1995)3396;\\
Phys. Rev. {\bf E53}(1996)2940.
\bibitem{Okano97}
K. Okano, L. Sch\"ulke, K. Yamagishi and B. Zheng, Nucl. Phys. 
{\bf B485 [FS]} (1997)727.
\bibitem{Luo98}
H.J. Luo, L. Sch\"ulke and B. Zheng, Phys. Rev. Lett. {\bf 81}(1998)180;\\
Phys. Rev. {\bf E57}(1998)1327.
\bibitem{Luo99}
H.J. Luo, L. Sch\"ulke and B. Zheng, 
accepted by  Mod. Phys. Lett. {\bf B}(cond-mat/9909325).
\bibitem{Ying98}
H.P. Ying, H.J. Luo, L. Sch\"ulke, and B. Zheng,  
Mod. Phys. Lett. {\bf B12}(1998)1237. 
\bibitem{Schue95}
L. Sch\"ulke, B. Zheng, Phys. Lett.{\bf A204}(1995)295.
\bibitem{Jas99}
A. Jaster, J. Mainville, L. Sch\"ulke and B. Zheng, 
J. Phys. A: Math. Gen. {\bf 32}(1999)1395. 
\bibitem{Janss89}
H.K. Janssen, B. Schaub, and B. Schmitmann, Z. Phys. {\bf B73}(1989)539.
\bibitem{Hoh77}
P.C. Hohenberg and B.I. Halperin, Rev. Mod. Phys. {\bf 49}(1977)435.
\bibitem{Bind92}
K. Binder and D.W. Heermann, {\it Monte Carlo Simulation in Statistical 
Physics} (Springer, Berlin, 1992).
\bibitem{Picco98}
M. Picco, e-print cond-mat/9802092.
\end{thebibliography}
\end{document}